\documentclass[twocolumn]{aastex62}

\renewcommand{\baselinestretch}{1.1}
\usepackage{textcomp}
\usepackage{gensymb}
\usepackage{epstopdf}
\usepackage{amsmath}
\usepackage{graphicx}
\usepackage{hyperref}
\usepackage{multirow}
\usepackage[nameinlink]{cleveref}
\usepackage[multiple]{footmisc}
\crefname{paragraph}{\S}{\S\S} 
\shorttitle{MOA-2009-BLG-319Lb} 
\shortauthors{S. K. Terry, et al.}

\begin{document}
\title{\textbf{MOA-2009-BLG-319Lb: A Sub-Saturn Planet Inside the Predicted Mass Desert}}

\author{Sean K. Terry}
\affiliation{Department of Physics, Catholic University of America, 620 Michigan Ave., N.E. Washington, DC 20064, USA}
\affiliation{Code 667, NASA Goddard Space Flight Center, Greenbelt, MD 20771, USA}

\author {Aparna Bhattacharya}
\affiliation{Code 667, NASA Goddard Space Flight Center, Greenbelt, MD 20771, USA}
\affiliation{Department of Astronomy, University of Maryland, College Park, MD 20742, USA}

\author{David P. Bennett}
\affiliation{Code 667, NASA Goddard Space Flight Center, Greenbelt, MD 20771, USA}
\affiliation{Department of Astronomy, University of Maryland, College Park, MD 20742, USA}

\author {Jean-Philippe Beaulieu}
\affiliation{School of Natural Sciences, University of Tasmania, Private Bag 37 Hobart, Tasmania, 70001, Australia}

\author{Naoki Koshimoto}
\affiliation{Code 667, NASA Goddard Space Flight Center, Greenbelt, MD 20771, USA}
\affiliation{Department of Astronomy, University of Maryland, College Park, MD 20742, USA}
\affiliation{Department of Astronomy, Graduate School of Science, The University of Tokyo, 7-3-1 Hongo, Bunkyo-ku, Tokyo 113-0033, Japan}

\author {Joshua W. Blackman}
\affiliation{School of Natural Sciences, University of Tasmania, Private Bag 37 Hobart, Tasmania, 70001, Australia}

\author {Ian A. Bond}
\affiliation{Institute of Mathematical and Natural Sciences, Massey University, Auckland 0745, New Zealand}

\author {Andrew A. Cole}
\affiliation{School of Natural Sciences, University of Tasmania, Private Bag 37 Hobart, Tasmania, 70001, Australia}

\author {Calen B. Henderson}
\affiliation{NASA Exoplanet Science Institute, IPAC/Caltech, Pasadena, CA 91125, USA}

\author{Jessica R. Lu}
\affiliation{Department of Astronomy, University of California Berkeley, Berkeley, CA 94701, USA}

\author {Jean Baptiste Marquette}
\affiliation{Laboratoire d'astrophysique de Bordeaux, Univ. Bordeaux, CNRS, B18N, alle Geoffroy Saint-Hilaire, 33615 Pessac, France}

\author{Cl\'ement Ranc}
\affiliation{Code 667, NASA Goddard Space Flight Center, Greenbelt, MD 20771, USA}


\author {Aikaterini Vandorou}
\affiliation{School of Natural Sciences, University of Tasmania, Private Bag 37 Hobart, Tasmania, 70001, Australia}

\correspondingauthor{S. K. Terry}
\email{41terry@cua.edu}

\begin{abstract}

\small \noindent We present\,\,an adaptive optics (AO) analysis of images from the Keck-II telescope NIRC2 instrument of the planetary microlensing event MOA-2009-BLG-319. The $\sim$10 year baseline between the event and the Keck observations allows the planetary host star to be detected at a separation of $66.5\pm 1.7\,$mas from the source star, consistent with the light curve model prediction. The combination of the host star brightness and light curve parameters yield host star and planet masses of $M_{\rm host} = 0.524 \pm 0.048M_{\sun}$ and $m_p = 67.3 \pm 6.2M_{\Earth}$ at a distance of $D_L = 7.1 \pm 0.7\,$kpc. The star-planet projected separation is  $2.03 \pm 0.21\,$AU. The planet-star mass ratio of this system, $q = (3.857 \pm 0.029)\times 10^{-4}$, places it in the predicted ``planet desert" at $10^{-4} < q < 4\times 10^{-4}$ according to the runaway gas accretion scenario of the core accretion theory. Seven of the 30 planets in the \citet{suzuki:2016a} sample fall in this mass ratio range, and this is the third with a measured host mass. All three of these host stars have masses of $0.5 \leq M_{\rm host}/M_{\sun}\leq 0.7$, which implies that this predicted mass ratio gap is filled with planets that have host stars within a factor of two of $1M_{\sun}$. This suggests that runaway gas accretion does not play a major role in determining giant planet masses for stars somewhat less massive than the Sun. Our analysis has been accomplished with a modified DAOPHOT code that has been designed to measure the brightness and positions of closely blended stars. This will aid in the development of the primary method that the \textit{Nancy Grace Roman Space Telescope} mission will use to determine the masses of microlens planets and their hosts.
\\
\\
\textit{Subject headings}: gravitational lensing: micro, planetary systems \\
\end{abstract}


\section{Introduction} \label{sec:intro}
\indent Gravitational microlensing has the unique ability to detect cold exoplanets beyond the 
snow line \citep{mao:1991a,gould:1992a} and down to Earth masses \citep{bennett:1996a}. So far microlensing has detected $\sim$100 planets at distances up to the Galactic Bulge. One drawback of this method is that for most light curves, only the mass-ratio of the lens system is measured, which leaves some physical parameters of the system significantly unconstrained. This results in large estimated uncertainties, particularly in the inferred stellar host and companion masses due to uncertain priors used in the standard Bayesian modeling approach.  One can mitigate this limitation by resolving the source and lens independently with high angular resolution imaging (i.e. \textit{Hubble Space Telescope} (HST), Keck AO, Subaru AO) several years after peak magnification, for which \cite{bennett:2006a,bennett:2007a} laid the theoretical groundwork. This high angular resolution imaging allows us to further constrain the lens-source separation, relative proper motion between the targets, and lens flux which can then be used with mass-luminosity relations
\citep{henry:1993a,henry:1999a,delfosse:2000a} to infer a direct mass for the host.\\
 \indent Several microlensing source and lens stars have now been measured with these techniques, beginning with OGLE-2005-BLG-169 \citep{bennett:2015a, batista:2015a}. These follow-up observations from Keck-II and HST confirmed, for the first time, the planetary interpretation from the light curve by verifying the lens-source relative proper motion as predicted by the original light curve measurement. The host star mass was precisely determined to be $0.69 \pm 0.02M_{\Sun}$, with a planetary companion of mass $14.1 \pm 0.9M_{\Earth}$.\\
\indent This current analysis is part of the NASA Keck Key Strategic Mission Support (KSMS) program,
``Development of the WFIRST Exoplanet Mass Measurement Method" \citep{bennett_KSMS}, which is a pathfinder project for 
the \textit{Nancy Grace Roman Space Telescope}  (formerly known as \textit{WFIRST}) \citep{spergel:2015a}. 
A large fraction of the \textit{Roman} Telescope observing time will be devoted to the \textit{Roman} Galactic Exoplanet Survey (RGES), which is a dedicated microlensing survey \citep{bennett:2002a,bennett_astro2010mpf,penny19,johnson:2020a} that will complement previous large statistical studies of transiting planets from the \textit{Kepler} telescope \citep{borucki:2011a} amongst others. The KSMS program has already measured the masses of several microlensing 
host stars and their planetary companions \citep{bhattacharya:2018a, vandorou:2019a, bennett:2020a}. Several more lens system mass measurements from the KSMS program are in preparation (Bhattacharya et al., in prep, Ranc et al., in prep, Blackman et al, in prep). A majority of the targets observed in this program were included in the statistical sample of \cite{suzuki:2016a,suzuki:2018a}, which shows a 
break and likely peak in the mass-ratio function for wide-orbit planets at about a Neptune mass. This study is the most complete statistical sample of microlensing planets to date, and the results are seemingly at odds with the runaway gas accretion scenario of the leading core accretion theory of planet formation \citep{lissauer93,pollack96}, which predicted a planet desert at sub-Saturn masses \citep{ida:2004a} for gas giants at wide orbits. \citet{suzuki:2018a} studied only the exoplanet mass ratio, $q$, 
so they could not determine if there was a gap over part of the host mass range. For example, since the core accretion theory was primarily developed with solar type host stars in mind, the gap expected from the runaway gas accretion scenario might
exist for solar-type stars, but be washed out with the low-mass M-dwarf hosts that are also included in the 
microlens sample. Mass measurements like the one presented in this paper can probe this possibility. \\
\indent This paper is organized as follows: Section \ref{sec:event} describes the original observations for 
MOA-2009-BLG-319. In Section \ref{sec:light-curve} we perform improved photometry of the light curve and present an updated analysis of the light curve. In Section \ref{sec:follow-up}, we describe the Keck adaptive optics (AO) follow-up analysis and a new MCMC routine for precise astrometry in Keck AO imaging. Section \ref{sec:prop-motion} details our lens-source relative proper motion measurements. Section \ref{sec:lens-properties} describes the lens system properties with new constraints from Keck high-resolution imaging. Finally, we discuss the results and conclude the paper in Section \ref{sec:conclusion}.


\section{Event MOA-2009-BLG-319 and New Photometry} \label{sec:event}
\begin{figure*}
\includegraphics[width=6.0in]{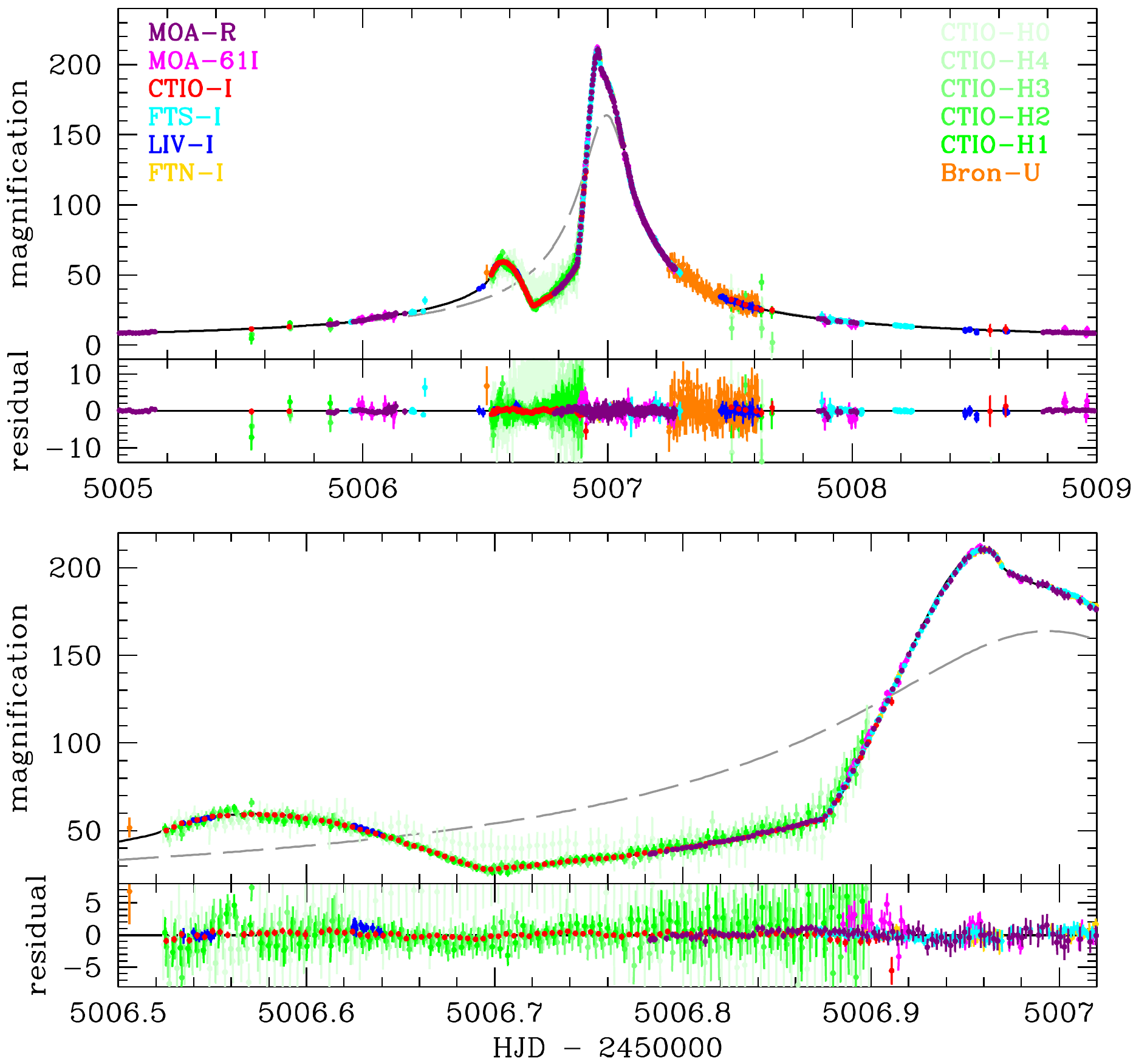}
\centering
\caption{\footnotesize Best fit planetary light curve model for MOA-2009-BLG-319 with the
data used for the analysis in this paper. Only the sparsely sampled $V$-band data is not shown. The CTIO-H0 through CTIO-H4 data are treated as independent data sets, shown in different shades of green. The data behind the figure is available in machine readable format. The data provided includes the Dan, Pal, and WISE I band measurements. All the data is presented in magnitudes units. \label{fig:lc}}
\end{figure*}

MOA-2009-BLG-319, located at RA $=$ 18:06:58.026, DEC $=$ -26:49:10.945 and Galactic coordinates ($l,b=(4.202, -3.014)$) was first alerted by the Microlensing Observations in Astrophysics 
(MOA; \citealt{bond01,sumi:2003a}) collaboration on 20th June 2009. MOA initially reported 
`low-level systematics' in their observations shortly after continuous monitoring began. This light curve feature turned out to be the first of several planetary caustic crossings throughout the duration of this high-magnification event. At the time of publication, MOA-2009-BLG-319 \citep{miyake:2011a} had the best sampled light curve of all observed microlensing events. \\
\indent Our photometry methods have improved since the \citet{miyake:2011a} analysis, so we have re-reduced the photometry for a number of the data sets. We have used the method of \citet{bond01,bond17} to reduce the data from the MOA-II telescope, the Mt.~John Observatory Boller and Chivens 0.61m telescope (operated by the MOA group), and the SMARTS telescope at CTIO. The MOA-II data were corrected for systematic errors due to chromatic differential refraction \citep{bennett12}. The SMARTS-CTIO data were previously reduced with DoPHOT \citep{dophot}, but the difference imaging photometry that we provide \citep{bond01,bond17} is well known to be a substantial improvement. New reductions are also needed to provide a Markov Chain Monte Carlo (MCMC) distribution to understand the distribution of models that are consistent with the data.\\
\indent While more than 20 data sets were used for the original paper, many of these do not actually constrain the light curve model. Therefore, we fit only to the following data sets: the MOA-II Red-band, the MOA 0.61m Boller and Chivens $V$ and $I$ band, SMARTS-CTIO $V$, $I$ and $H$ band, the Robonet Faulkes telescope (North and South) $I$ band, the Liverpool telescope $I$ band, and the Bronberg Observatory unfiltered data. Figure \ref{fig:lc} shows the best fit model with the data used in this paper, except for the sparsely sampled $V$ band data. The CTIO data were taken with the ANDICAM instrument of the SMARTS-CTIO telescope, which takes optical and infrared data simultaneously. The infrared data from this telescope is known to occasionally display systematic errors between images taken at the five different dither positions, that are apparently due to sub-pixel scale sensitivity variations \citep{dong-moa400}. Therefore, we treat the data from these different dither positions as independent
data sets, shown in different shades of green in Figure~\ref{fig:lc} as CTIO-H0 through CTIO-H4.

\section{New Light Curve Model} \label{sec:light-curve}
The light-curve modeling follows the image-centered ray shooting method of \cite{bennett:1996a} and \cite{bennett:2010a}. Figure~\ref{fig:lc} shows our best fit planetary model for this event and Table \ref{tab:lcpar} shows the parameters of our best-fit model, as well as the MCMC averages of models 
consistent with the data. These are also compared to the distribution from the original study of \cite{miyake:2011a}. 
\\
\indent A follow-up light curve analysis by \citet{shin:2015a} considered two-planet models for MOA-2009-BLG-319
and a number of other planetary microlensing events, and their analysis found a significant
$\chi^2$ improvement, $\Delta\chi^2 > 100$, for their best two planet model for this event. However, this analysis
was incomplete, as they did not consider other triple-lens models for this event. The analysis of planetary
microlensing event OGLE-2007-BLG-349 indicates that circumbinary models can describe deviations that
are also consistent with two-planet models \citep{bennett16}, and there can also be degeneracies between
circumbinary planet models and circumstellar planet models in binary systems \citep{gould-ogle341}.
We will not consider these triple lens models further in this paper, as the analysis of these triple lens models
is not complete. We should note, however, that if the two-planet model is correct, then the conclusions
of this paper will be unchanged except that there will be an additional, lower-mass planet. Also, these
triple lens models are relevant for the consideration of a microlensing parallax signal.
While the MOA-2009-BLG-319 Einstein radius crossing time is too short to expect a microlensing parallax signal due to the orbital motion of the Earth, the dense coverage of the light curve peak by widely separated observatories suggests the possibility of a terrestrial microlensing parallax signal \citep{hardy95,holz96,gould09}, as pointed out by \citet{miyake:2011a}. However, the triple lens models will effect the same part of the light curve. Thus, it would not be useful to investigate any microlensing parallax solution without also considering a third lens mass.\\
\indent In order to determine the source radius, we need to determine the extinction corrected source magnitude and color. \citet{miyake:2011a} used the SMARTS-CTIO $V$ and $I$ band data for this. However, these SMARTS-CTIO data were reduced with DoPHOT, and this has occasionally led to magnitude and color measurements that led to spurious conclusions about the properties of planetary microlens systems \citep{bennett17}. This is why it was necessary to use the difference imaging code and calibration method of \citet{bond17} for this reanalysis of the SMARTS-CTIO $V$ and $I$ band data. Also, predicted properties of the bulge red clump giant stars that are used to determine the extinction have changed since the \citet{miyake:2011a} analysis. We have calibrated the SMARTS-CTIO $V$ and $I$ band data to the OGLE-III catalog \citep{ogle3-phot}, and then we located the red clump centroid at $V_{\rm rc} - I_{\rm rc} = 1.98$, $I_{\rm rc} =15.44$, following the method of \citet{bennett:2010b}. Using the bulge red clump giant magnitude, color, and distance from \citet{nataf:2013a}, we find $I$ and $V$ band extinction of $A_I = 1.116$ and $A_V = 2.036$. Using the source magnitudes from Table~\ref{tab:lcpar}, we find extinction corrected magnitudes of $I_{S0} = 18.878 \pm 0.069$ and $V_{S0} = 19.678 \pm 0.069$. This allows us to use the surface brightness relation from the analysis of 
\citet{boyajian:2014a}, but we use the following custom formula \citep{aparna16} using stars spanning the range in colors that are relevant for microlensing events:
\begin{equation}
    \textrm{log}({2\theta_{*}}) = 0.5014 + 0.4197(V_{S0} - I_{S0}) - 0.2 I_{S0}
\end{equation}

\noindent This yields $\theta_{*} = 0.576 \pm 0.077 \mu$as, which is smaller than the \cite{miyake:2011a} value of $\theta_{*} = 0.66 \pm 0.06 \mu$as. Our measurement is consistent with the $\mu_{\rm rel}$ measurement from Keck. This difference from the value that \cite{miyake:2011a} find is due in part to the combination of the error in magnitude from DoPHOT and an improved knowledge of the red clump from \cite{nataf:2013a} as described earlier.

\begin{deluxetable*}{lcccr}[t]
\deluxetablecaption{Best Fit MOA-2009-BLG-319L Model Parameters\label{tab:lcpar}}
\tablecolumns{5}
\setlength{\tabcolsep}{26.5pt}
\tablewidth{\columnwidth}
\tablehead{
\colhead{Parameter} &
\colhead{Units} & \colhead{Value} & \colhead{MCMC Averages} & \colhead{\hspace{6mm}Miyake+2011}
}
\startdata
$t_E$ & days & $16.762$ & $16.72 \pm 0.10$ & $16.56 \pm 0.08$\\
$t_{0}$ & HJD$'$ & $5006.9951$ & $5006.9952 \pm 0.0008$ & $5006.995 \pm 0.001$\\
$u_0$ & {} & $-0.006103$ & $-0.0061 \pm 0.0004$ & $-0.0062 \pm 0.0003$\\
$s$ & {} & $0.97564$ & $0.9756 \pm 0.0001$ & $0.975 \pm 0.001$\\
$\alpha$ & radians & $-2.62995$ & $-2.6299 \pm 0.0007$ & $-2.629 \pm 0.001$\\
$q \times 10^4$ & {} & $3.8463$ & $3.856 \pm 0.029$ & $3.95 \pm 0.02$\\
$t_*{}$ & {days} & $0.03186$ & $0.0319 \pm 0.0006$ & $0.0320 \pm 0.0033$\\
$I_s{}$ & {} & $19.994$ & $19.992 \pm 0.007$ & $19.78 \pm 0.07$\\
$V_s{}$ & {} & $21.714$ & $21.712 \pm 0.007$ & $21.52 \pm 0.09$\\
$\chi^2 / \textrm{dof}$ & {} & $10746.24/10805$ & {} & {}
\enddata
\tablenotetext{}{\footnotesize{\textbf{Notes.} HJD$'$ = HJD$-2450000$. \cite{miyake:2011a} values are for their best-fit $u_{0} < 0$ solution without parallax. We have performed a change of coordinate for $\alpha$ reported in \cite{miyake:2011a}} by $\pi \rightarrow \pi - \alpha$, based on the choice of `mass one' for the planet.}
\end{deluxetable*}

\indent To measure our new lens system parameters, we sum over the 
MCMC results using a Galactic model \citep{bennett:2014a} with weights for the microlensing rate and our $\mu_{\textrm{rel,H}}$ value from Keck (described in Section \ref{sec:prop-motion}). We constrain the possible source distances to follow the weighted distribution from the microlensing event rate in our Galactic model, which results in a best-fit source distance of $D_{S} = 8.25 \pm 0.86$ kpc. These new light curve modeling results produce smaller best-fit values for the mass ratio, $q$, and angular Einstein radius $\theta_{\textrm{E}}$, and larger $t_{E}$ value as can be seen in table \ref{tab:lcpar}. This difference is due to the new de-trended MOA-R and CTIO difference imaging photometry.\\ 
\indent Since we do not have a measurement of the microlensing parallax $\pi_E$, we use the Keck lens flux and mass-luminosity relations \citep{henry:1993a,henry:1999a,delfosse:2000a} in order to constrain the lens distance. The extinction in the foreground of the lens is calculated assuming a dust scale height of $h_{\textrm{dust}} = 0.10\pm 0.02\,$kpc. 


\begin{figure*}
\includegraphics[width=\linewidth]{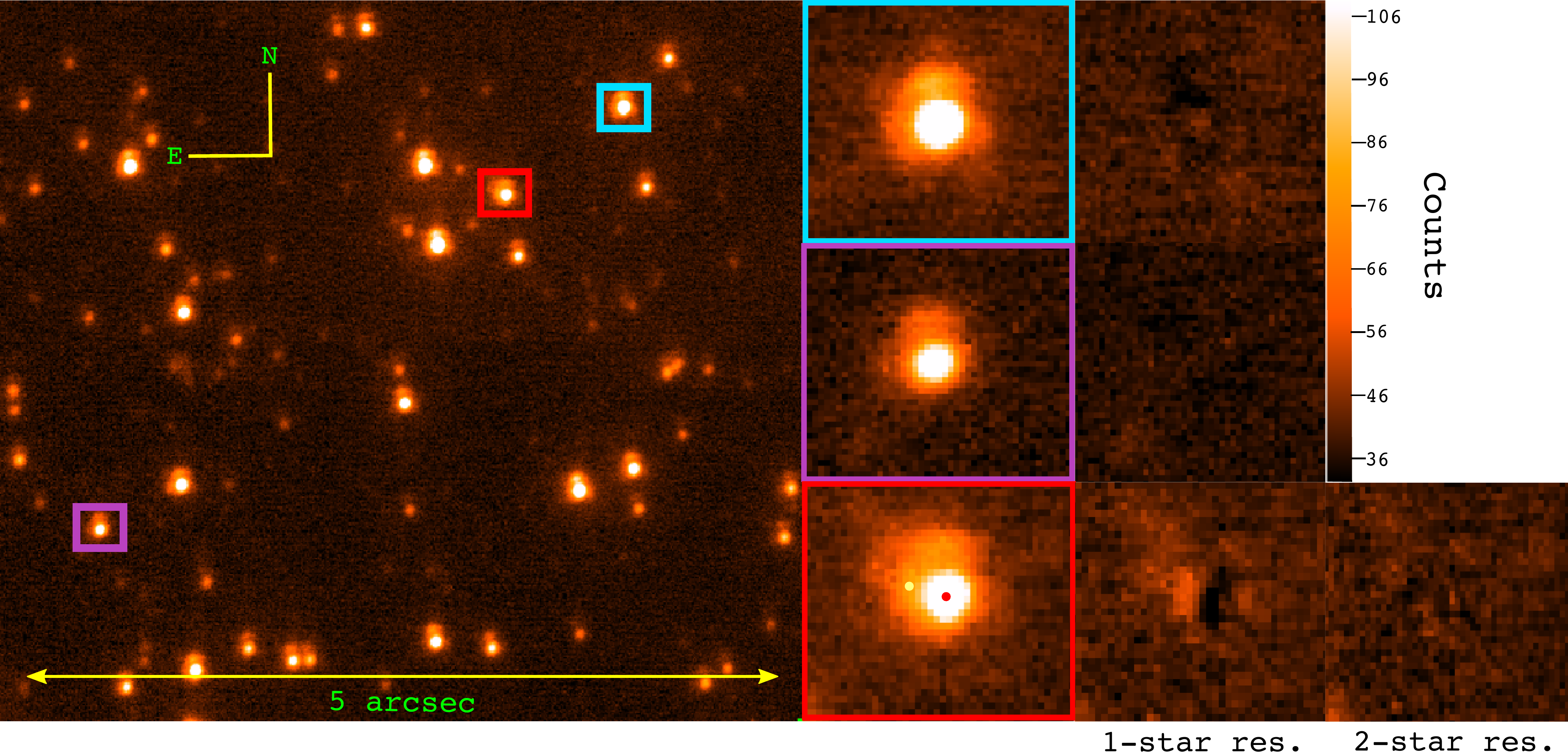}
\centering
\caption{\footnotesize \textit{Left Panel}: Co-added sum of 9 60-sec NIRC2 $K$ band narrow camera images from 2019. \textit{Cyan, purple panels}: closeup of single stars in the frame, with 1-star PSF residuals plotted next to each. \textit{Red panel}: closeup of MOA-2009-BLG-319 showing center position of the source (red point) and lens (yellow point), with 1-star and 2-star PSF residuals respectively. The color-bar refers to the PSF residual images only. \label{fig:oct-panel}}
\end{figure*}

\section{Keck Follow-up and Analysis} \label{sec:follow-up}
The target MOA-2009-BLG-319 was observed with the NIRC2 instrument on Keck-II in the $H$ and $K_{short}$ (hereafter $K$) on May 25, 2018 and $K$ band on May 28, 2019. The 2018 $K$ band data have a point-spread function (PSF) full-width 
half-max (FWHM) of $\sim$70 mas. The 2018 $K$ band data have somewhat poorer quality than the 2019 $K$ band data, and the 2018 $H$ band data is even more problematic, with a larger PSF (FWHM$\sim$120 mas). 
In section \ref{kband-2018}, we discuss the analysis of the 2018 $K$ band data, and in section \ref{hband-analysis} we test the limits of our detection capabilities with the very marginal 2018 $H$ band signal.\\
\indent For the 2018 and 2019 observations, both the NIRC2 wide and narrow cameras were used. The pixel scales for the wide and narrow cameras are 39.69 mas/pixel and 9.942 mas/pixel, respectively. All of the images were taken using the Keck-II laser guide star adaptive optics system. \\
\indent As we discuss below in Sections~\ref{kband-2018} and \ref{hband-analysis}, our highest precision measurements come from the 2019 data, so we will focus on the analysis of that data. For the 2019 data, a co-add of 9 dithered wide camera images were used for photometric calibration to images from the Vista Variables in the Via Lactea (VVV) survey \citep{minniti:2010a} following the procedure of \cite{beaulieu:2018a}. The wide camera images were flat-field and dark current corrected using standard methods, and stacked using the SWarp software \citep{bertin:2010a}. We performed astrometry and photometry on the co-added wide camera image using SExtractor \citep{bertin:1996a}, and subsequently calibrated the narrow camera images to the wide camera image by matching two dozen bright isolated stars in the frames. This calibration analysis results in uncertainties of 0.06 magnitudes.\\
\indent For the 2019 $K$ band narrow data, we combined 30 flat-field frames, 10 dark frames, and 15 sky frames for calibrating our science images. Following the methods of \cite{service:2016a} and \cite{yelda:2010a}, we then combined 9 $K$ band narrow camera science frames with an integration time of 60 seconds per frame. The combined frame can be seen on the left panel of figure \ref{fig:oct-panel}, which has a PSF full width at half maximum (FWHM) of $\sim$73mas. The reduction of the 2018 $H$ band data follows the same pipeline as the $K$ band described in this section.\\
\indent Lastly, there were 10 $K$ band images of the target taken on July 26, 2015 with the NIRC2 narrow camera that were combined to make one co-added science frame. There were no sky frames taken for the 2015 data, which contributes to the lower signal-to-noise seen in this data. The much smaller lens-source separation at the time of these images also implies that our lens-source relative proper motion, $\mu_{\textrm{rel,H}}$, and lens brightness measurements, will be less precise than the later images. Even with this lower signal data, a careful DAOPHOT reduction successfully detects the lens. Further details of the 2015 analysis are given in section \ref{2015-analysis}. The main benefit of these early images is that they allow us to verify the identification of the lens star, by showing that it is moving away from the source at a rate
consistent with the occurrence of the microlensing event in June, 2009.

\subsection{PSF Fitting Photometry }\label{sec:photometry}
Because the two stars in the blend have a separation in 2019 of $\sim$FWHM, it is necessary to use a PSF fitting routine to measure both targets independently. Following the methods of \cite{bhattacharya:2018a} and references therein, we use the photometry routine DAOPHOT-II \citep{stetson:1987a} to generate and fit an empirical PSF to the source+lens blend. 
The AO corrections for observations of our Galactic bulge fields using the instruments currently on the Keck telescope
generally deliver imperfect AO corrections with Strehl ratios $< 0.5$, and often the Strehl ratios are significantly smaller
than 0.5. Thus, the PSFs delivered by the AO system can have a wide variety of shapes. The DAOPHOT package
has proven to be quite successful in modeling oddly shaped PSFs delivered by the Keck AO system \citep{bennett:2010b}.
An alternative method has also been presented by \citet{vandorou:2019a}, that is probably competitive with DAOPHOT. DAOPHOT's sophisticated semi-empirical PSF is important for our observations of MOA-2009-BLG-319 since the PSF has a prominent wing to the North that has a similar amplitude to the flux ratio to the companion star to the MOA-2009-BLG-319S source star that we interpret to be the lens star (MOA-2009-BLG-319L).\\
\indent The first pass of DAOPHOT does not detect the lens, but instead produces a clear feature to the East in the residual image which can be seen in the lower-right panel (labeled ``\texttt{1-star res.}") of figure \ref{fig:oct-panel}. The target is the only stellar image that has an extension in this direction, and this feature represents the position of the fainter lens star. The cyan and purple panels in figure \ref{fig:oct-panel} show reference stars in the frame with similar brightness as the target that also exhibit the PSF extension to the North. This extension is accurately modeled by the DAOPHOT single-star PSF model as can be seen by the featureless residuals to the right of each reference star. The color-bar on the right represents the pixel counts for the residual images only. The lens also has a separation consistent with that predicted by \cite{miyake:2011a}, this separation is described further in section \ref{sec:prop-motion}.\\
\indent Fitting a two-star PSF to the target and re-running DAOPHOT produces a nearly featureless residual, shown in the lower-right panel (labeled ``\texttt{2-star res.}") of figure \ref{fig:oct-panel}. Table \ref{table:dual-phot} shows the calibrated magnitudes for the two stars of $K_{S}=18.12 \pm 0.05$ and $K_{L}=19.98 \pm 0.09$. The uncertainties are derived from the ``jackknife method" described in Section~\ref{sec:jackknife}. Using the VVV extinction calculator \citep{gonzalez:2011a} and the \cite{nishiyama:2009a} extinction law, we find a $K$ band extinction of $A_K = 0.13 \pm 0.05$. From our re-analysis of the light curve modeling (Section \ref{sec:light-curve}), we find a source color of $V_S - I_S = 1.72$, which leads to an extinction-corrected color of $V_{S0} - I_{S0} = 0.80$. We use the color-color relations of \cite{kenyon:1995a} and the I-band magnitude, $I_S = 19.994$ to predict a source $K$ band magnitude of $K_S = 18.15$. The fit source brightness is fainter than our measured source brightness by less than 1$\sigma$, thus we conclude that there is virtually no evidence of additional flux from a companion to the source.

\begin{deluxetable}{lcr}[!h]
\deluxetablecaption{2019 Dual-Star PSF Photometry\label{table:dual-phot}}
\tablecolumns{3}
\setlength{\tabcolsep}{10.0pt}
\tablewidth{\linewidth}
\tablehead{
\colhead{\hspace{-15mm}Star} &
\colhead{Passband} & \colhead{\hspace{19mm}Magnitude}
}
\startdata
Lens & Keck $K$ & $19.98 \pm 0.09$\\
Source & Keck $K$ & $18.12 \pm 0.05$\\
Source $+$ Lens & Keck $K$ & $17.94 \pm 0.06$\\
\enddata
\tablenotetext{}{\footnotesize{\textbf{Note}. Magnitudes are calibrated to the VVV scale, as described in section \ref{sec:follow-up}.}}
\end{deluxetable}


The standard version of DAOPHOT has some drawbacks for our problem of studying the closely 
blended images of microlens source and lens stars. First, we want to be able to study cases where the
detection of the lens star may be marginal, as well as cases we can only obtain an upper limit on the lens brightness as a function of the lens-lens source separation. Thus, it would be useful to have a method that will produce a probability distribution of all possible source plus lens configurations that are consistent with the data. The standard version of DAOPHOT, on the other hand, is programmed to avoid including false detections in its output star list, so it may reject some of the more marginal lens detections. Of course, because of the microlensing event, we know that another star is there, although it might be quite faint (Blackman et al., in preparation). Also, as \citet{bennett:2007a} have shown, constraints on the source brightness and/or lens-source separation from the light curve models can often significantly reduce the uncertainties on parameters, such as the lens brightness, that are not significantly constrained by the light curve data. Thus, it will be useful to be able to apply these constraints inside of DAOPHOT in order to get the most precise possible measurement of the lens star properties. \\
\indent Finally, DAOPHOT does not report error bars on the star positions, which are critical for our science. \citet{king:1983a} did publish a formula that can be used to estimate position error bars based on the photometry error bars, but this formula is problematic for our situation of highly blended stellar images. We can address these issues by modifying the standard version of DAOPHOT and adding a routine that uses the Markov Chain Monte Carlo (MCMC) method to determine the distribution of source and lens star magnitudes and positions that are consistent with the data, as we explain in the next subsection.

\subsubsection{Development of a MCMC Routine for DAOPHOT}\label{sec:mcmc}

\begin{figure}
\includegraphics[width=\linewidth]{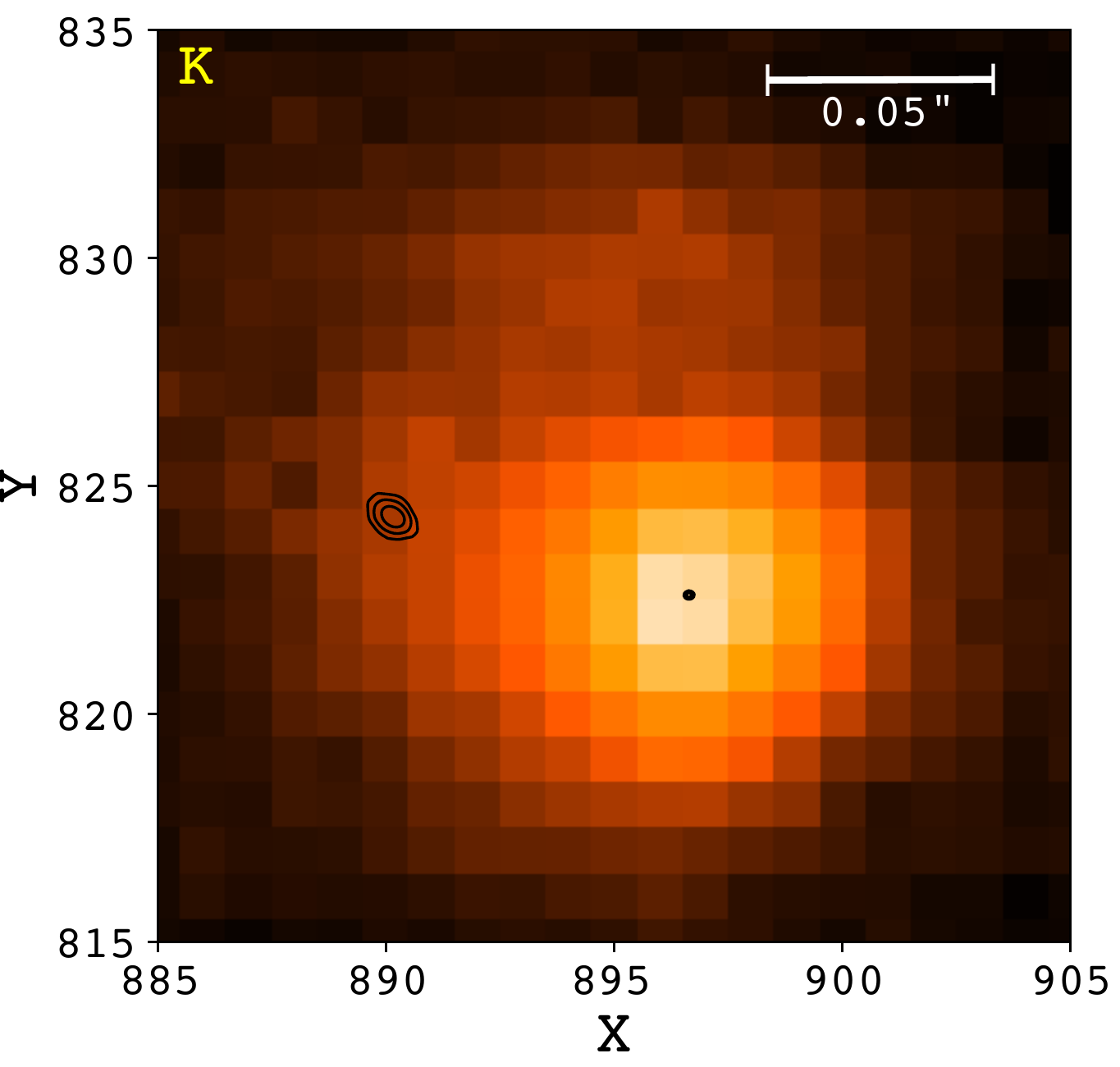}
\centering
\caption{\footnotesize Best fit MCMC contours (68.3\%, 99.5\%, 99.7\%) for the source and lens positions respectively, over-plotted on the $K$ band image of the target. The lens contributes $\sim$15\% of flux to the total blend. \label{fig:contour}}
\end{figure}

\indent We begin the MCMC routine by using the DAOPHOT empirical PSF that was described in the previous section. This PSF model is then permitted to step across the fitting box encompassing the blended targets, with a fitting box radius of $\sim$1.5 FWHM. For the dual-star version of the MCMC routine, there are six total parameters that are simultaneously fit: the x and y pixel location for each star ($x_1,y_1,x_2,y_2$), the total flux ($f_{T}$), and the flux ratio between the stars ($f_{R}$). For each step in the chain, a $\chi^2$ value for the fit is measured and recorded. The routine then takes a random step in any direction (and flux), makes the same measurements and compares the new $\chi^2$ to the previous. If the new value is smaller than the previous, the 6-parameter fit is recorded and the routine continues. However, If the new value is larger than the previous, a weighted proposal probability distribution is calculated. If this weighted probability is less than a randomly generated probability (between $0-1$), then the decision is reversed and the original candidate value is accepted. If the weighted probability is greater still, the candidate is rejected and the iteration moves forward with a new candidate. This procedure follows the standard Metropolis-Hastings method. Once the routine has converged, the best-fit parameters are recorded.\\
\indent For a dual-star model, we calculate the flux distribution following \cite{bhattacharya:2017a}:
\begin{equation}\label{eq:3}
    f_{T} = f_{1}\psi(i - x_1,j - y_1) + (1-f_{1})\psi(i - x_2,j - y_2),
\end{equation}
\noindent where $f_1$ is the source flux contribution to the total flux, $1 - f_1$ is the lens flux contribution, and $\psi$ is the 2-dimensional PSF model. The values $x_1, y_1, x_2, y_2$ are the initial pixel positions for the source and lens as described earlier in this section, and the indices $i$ and $j$ are the trial pixel positions for a given iteration. The $\chi^2$ minimization routine described above computes the minimum value of the six-parameter fit as follows:

\begin{equation}\label{eq:chi2}
\begin{split}
    \chi^2 & = \sum_{i,j}[\frac{1}{\sigma}\{P_{i,j} - s_{*} - f_{1}\psi(i - x_{1},j - y_{1}) \\
    & - (1 - f_{1})\psi(i - x_{2},j - y_{2})\}]^2,
\end{split}
\end{equation}

\noindent where $P_{i,j}$ is the intensity at pixel location $i,j$, $\sigma$ is the uncertainty in pixel intensity, and $s_{*}$ is the background flux. The MCMC chains are used as a probability distribution that we use to determine the normalized errors on the best-fit MCMC results in Tables \ref{tab:mcmc-jackknife} and \ref{tab:h-vs-k}.\\
\indent The standard version of DAOPHOT employs the Newton-Raphson method \citep{press:1986a} for fitting the positions of the two blended stars. The two-star routines were run with both the Newton-Raphson and MCMC methods, producing nearly identical results. The residual images for the reference stars shown in Figure~\ref{fig:oct-panel} are the residuals from the Newton-Raphson analysis of standard DAOPHOT. The residual images for the target shown in the same figure are from the MCMC analysis. Figure \ref{fig:contour} shows the 1$\sigma$, 2$\sigma$, and 3$\sigma$ contour intervals for the best-fit MCMC source and lens positions, over-plotted on the stellar image. The best-fit parameters from the MCMC routine and their respective error bars are listed in table \ref{tab:mcmc-jackknife}, along with the error bars from the jackknife method as discussed in subsection~\ref{sec:jackknife}. The lens-source separation measurement with our MCMC routine is within 1$\sigma$ of the result from standard DAOPHOT.\\
\indent The routine also has the functionality to fit the simpler case of a single star. This single-star MCMC fitting was performed on the source+lens blend and produced the residual seen in figure \ref{fig:oct-panel} (``\texttt{1-star res.}"). The two-star MCMC run produces a better fit as expected, with a $\chi^2$ improvement of $\Delta\chi^2 = 1313.0$ over the single-star fit. The residual image that was created using the MCMC best-fit two-star values is nearly 
featureless and produced the residual shown in the lower right panel (``\texttt{2-star res.}") of figure \ref{fig:oct-panel}.\\

\begin{deluxetable*}{@{\extracolsep{4pt}}llrlrlrlrlr}
\tablecaption{DAOPHOT MCMC and Jackknife Best Fit Results \label{tab:mcmc-jackknife}}
\setlength{\tabcolsep}{5.0pt}
\tablewidth{\columnwidth}
\tablehead
{
\colhead{}&
  \multicolumn{2}{c}{$2015$ $K$ band}&
  \multicolumn{2}{c}{$2018$ $K$ band}&
  \multicolumn{2}{c}{$2019$ $K$ band} \\
\cline{2-3} \cline{4-5} \cline{6-7}
\colhead{\hspace{-2.0cm}Parameter} & \colhead{MCMC}& 
\colhead{Jackknife} & \colhead{MCMC} & \colhead{Jackknife}&
\colhead{MCMC} & \colhead{Jackknife}
}
\startdata
$\mu_{\textrm{rel,HE}}$ (mas/yr) & $6.134 \pm 1.281$ & $6.970 \pm 2.187$ & $7.172 \pm 0.472$ & $6.669 \pm 0.311$ & $6.482 \pm 0.167$ & $6.405 \pm 0.072$\\
$\mu_{\textrm{rel,HN}}$ (mas/yr) & \hspace{-0.25cm}$-1.351 \pm 0.775$ & $-0.555 \pm 2.034$ & $0.656 \pm 0.290$ & $0.568 \pm 0.309$ & $1.684 \pm 0.158$ & $1.788 \pm 0.145$\\
Lens Flux/Source Flux & $0.129 \pm 0.069$ & $0.158 \pm 0.053$ & $0.176 \pm 0.008$ & $0.176 \pm 0.047$ & $0.176 \pm 0.007$ & $0.180 \pm 0.014$\\
\enddata
\end{deluxetable*}

\subsubsection{Error Bars with the Jackknife method}\label{sec:jackknife}
While the MCMC method is a powerful tool for studying the range of model parameters that are consistent with an image, there is another source of uncertainty that we must consider for our analysis of Keck adaptive optics images. It is standard practice to analyze combinations of multiple dithered infrared images in order to remove some of the instrumental artifacts from these images. However, the adaptive optics images have imperfect corrections to the optical effects of the atmosphere. The quality of the adaptive optics correction is often characterized by the Strehl ratio, which is the ratio of the brightness at the peak of a stellar PSF, to the peak that would be obtained due only to diffraction. In moderately good observing conditions, like the conditions for our 2019 $K$ band observations of MOA-2009-BLG-319, we typically have Strehl ratios in the range 0.2-0.4. In $H$ band, the Strehl ratios are worse, typically 0.1-0.2, although these images can have PSF FWHM values as good or better than the $K$ band images with better Strehl ratios. Thus, greatly improved angular resolution given by these adaptive optics systems yields images that are far from perfect. Significant PSF distortions remain in the Keck AO images, and these distortions vary from image to image, and it is also likely that there is some variation across each image. Because of this, we measure the PSF with stars close to the target in our analysis, but we must also consider the effect of the variations between images.\\
\indent The uncertainty due to the variations between images can be addressed by the jackknife method \citep{quenouille49,quenouille56,tukey58}. Our implementation of this method is discussed in more detail by
\cite{bhattacharya:2020a}. To analyze a collection of $N$ dithered images, we create $N$ different combinations of $N-1$ images, with each image missing from only one of these combinations. The error bars for each parameter are then given by $\sqrt{N-1}$ times the RMS of the best fit parameters from each of these $N$ combinations of $N-1$ images. Table~\ref{tab:mcmc-jackknife} compares the error bars computed by the MCMC method to the error bars computed by the jackknife method. We chose to use the jackknife uncertainties because they include the uncertainties due to the PSF variations in the individual images.

\subsection{2018 $K$ band Analysis}\label{kband-2018}
In addition to the 2019 $K$ band data discussed in detail above, we also obtained a set of 13 30-second exposure NIRC2 narrow camera images on May 25, 2018. A total of 20 calibration frames were used for flat-fielding, dark subtraction, and sky subtraction.\\
\indent The 2018 $K$ band images have a PSF FWHM similar to the 2019 $K$ band images, although the PSF appears to be slightly elongated in the East-West direction instead of having the extended wing to the North, like the 2019 $K$ band images. This is a complication because the lens star is located toward the East, but the more serious issue is that these images are much noisier. They have been taken through $\sim 0.7$ mag of extinction due to cirrus clouds, and there appears to have been a substantial amount of moonlight reflected off the clouds. This generated a much higher background and probably prevented the sky subtraction from removing some systematic errors. \\
\indent We reduced these data with the same procedures used for the 2019 $K$ band data described above, and the results were very similar to the 2019 $K$ band results. However, as shown in Table~\ref{tab:mcmc-jackknife}, the error bars from the jackknife method were significantly larger than for the 2019 $K$ band data, particularly for the lens/source flux ratio and the $\mu_{\rm rel,HE}$ component of the relative proper motion. Therefore, we use the 2019 $K$ band data for our constraints on the properties of the lens system, although the results with the weighted sum of the 2018 and 2019 $K$ band data are indistinguishable.

\begin{figure*}[!ht]
\includegraphics[width=\linewidth]{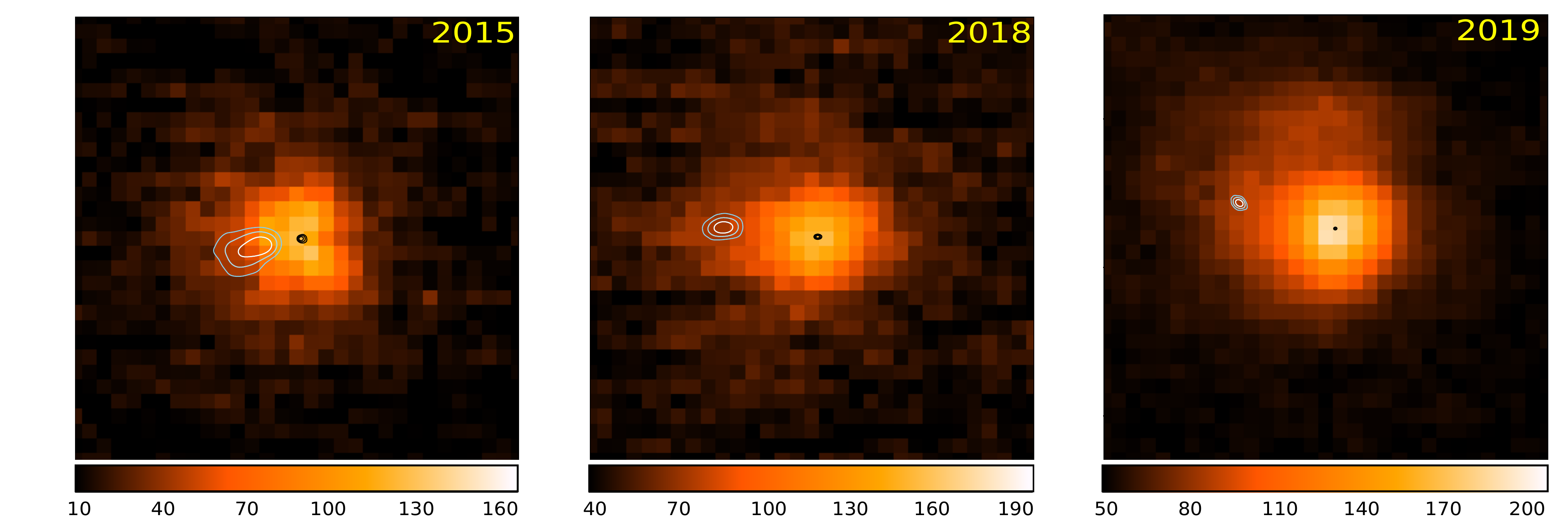}
\centering
\caption{\footnotesize The best fit MCMC contours (68.3\%, 99.5\%, 99.7\%) for the source and lens positions are shown over-plotted on the 0.3"$\times$0.3" $K$ band images from 2015 (left), 2018 (middle), and 2019 (right). The color bar refers to the pixel intensity. North is up and East is left in all panels. This series of data clearly show that the lens and source are separating from each other. While the MCMC calculations provide enough resolution to calculate contours, they can often be underestimated because they exclude any effects of PSF variations between images. \label{fig:2015-2018-2019}}
\end{figure*}

\subsection{2018 $H$ band Analysis with Lens-Source Separation Constraint}\label{hband-analysis}
The clear lens detection from the 2019 $K$ band data allows us to carefully test the capabilities of our observations and analysis on a data set with a marginal detection (i.e. the $H$ band data for MOA-2009-BLG-319). Our initial reduction of this data with standard \texttt{DAOPHOT} did not detect the lens or show an obvious feature in the best-fit single-star residual to indicate the presence of the lens star. In addition, our first attempts at two-star fits with the MCMC version of \texttt{DAOPHOT} also did not successfully converge on a lens location. Following the methods described in \cite{bhattacharya:2017a}, we implemented a separation constraint to our MCMC analysis based on the known $\mu_{\rm rel}$ from our light curve re-analysis. While we could also constrain the 2018 separation based on our 2019 $K$ band lens-source separation measurement, our goal is to show the reliability of a marginal detection with MCMC on future targets that do not have any such better data. With this lens-source separation constraint, along with a renormalization of the pixel errors such that the best-fit $\chi^2/\rm{d.o.f} \simeq 1$, the MCMC converged on a solution for the lens location of $57.5 \pm 2.4$ mas to the NE of the source, consistent with the 2019 data. The renormalization factor for our $H$ band analysis was 0.256, and the total number of fitted pixels was 2304. Finally, we test the stability of the PSF model by calculating the total $\chi^2$ of the pixels from a radius of one pixel from the center of the bright source, to a radius the size of the fitting box. We find a relatively smooth distribution in $\chi^2$/pix space, which indicates a stable PSF model.\\
\indent We subsequently re-ran the MCMC routine with the separation constraint and renormalized errors, and our best-fit results show that the lens is detected, albeit with less confidence than the $K$ band result. The best-fit results for the $H$ band are shown in Table \ref{tab:h-vs-k}. One drawback we find during this marginal detection test is that the best-fit lens-source flux ratio is not consistent with the 2019 result. The contrast should be somewhat lower in $H$ band since the lens is redder than the source, however the $H$ band results are more than $10\sigma$ lower than $K$ band.

\subsection{2015 $K$ band Analysis} \label{2015-analysis}
We performed a DAOPHOT analysis of the 2015 $K$ band data, similar to that of the previous reductions. The PSF FWHM for this data is approximately 75 mas, which means the lens-source separation is $\sim$0.53$\times$ FWHM at the time of the 2015 data approximately 6.09 years after $t_0$. The lens-source relative proper motion, $\mu_{\rm rel}$, and flux ratio for the 2015 data is given in Table \ref{tab:h-vs-k}. The East and North component of the Heliocentric relative proper motion from the Jackknife method is consistent with both the 2018 and 2019 $K$ band data. Figure \ref{fig:2015-2018-2019} shows the best fit MCMC contours for the source and lens positions for each epoch, with the $K$ band image over-plotted. The color bar refers to the pixel intensities in each frame. It is clear from these results that we are in fact measuring the lens and source moving away from one another.\\
\indent The main contribution of these 2015 images is not to increase the precision of our $\mu_{\textrm{rel,H}}$ measurements. Instead it serves to confirm our identification of the lens star. As can be seen in Table~\ref{tab:mcmc-jackknife}, the
$\mu_{\textrm{rel,H}}$ measurements from the 2015 images are consistent with the much more precise 2019 measurements. In particular, the $\mu_{\textrm{rel,HE}}$ value is within 0.25$\sigma$ of the 2019 value, and the $\mu_{\textrm{rel,HN}}$ value is within 1.2$\sigma$ of the 2019 value (using the jackknife error bars). \\
\indent The observed motion between 2015 and 2019 rules out a possible companion to the source star as the source of the flux that we attribute to the lens star. The implied velocity is much too large for the star to be bound to the source. An unrelated star in the bulge would have to mimic the proper motion of the lens star, and the probability of this is $\lesssim 10^{-4}$ according to an analysis using the method of \cite{koshimoto:2020a}. There is also the possibility that we have detected the combination of the flux of the planetary host and a binary companion to the host star. The \cite{koshimoto:2020a} analysis predicts a probability of 1.9\% for this possibility, but this does not include a complete analysis of the triple lens modeling for this event. There is a weak signal that could be due to an additional planet \citep{shin:2015a} or an additional star, but this will be investigated in detail in a subsequent paper.

\begin{deluxetable*}{lcccc}[!htp]
\tablecaption{Best Fit MCMC Results for Relative Proper Motion and Flux Ratio\label{tab:h-vs-k}}
\tablecolumns{5}
\setlength{\tabcolsep}{23.5pt}
\tablewidth{\columnwidth}
\tablehead{
\colhead{\hspace{-1.8cm}Parameter} & \colhead{2015 $K$} & \colhead{2018 $H$} & \colhead{2018 $K$} & \colhead{2019 $K$}
}
\startdata
$\mu_{\textrm{rel,HE}}$ (mas/yr) & $6.134 \pm 1.281$ & $6.183 \pm 0.449$ & $7.172 \pm 0.472$ & $6.482 \pm 0.167$\\
$\mu_{\textrm{rel,HN}}$ (mas/yr) & $-1.351 \pm 0.775$ & $1.823 \pm 0.889$ & $0.656 \pm 0.290$ & $1.684 \pm 0.158$\\
Lens Flux/Source Flux & $0.129 \pm 0.069$ & $0.034 \pm 0.009$ & $0.176 \pm 0.008$ & $0.175 \pm 0.007$\\
\enddata
\tablenotetext{}{\footnotesize{\textbf{Notes}. 2018 $H$ band lens-source flux ratio is unreliable, as described in section \ref{hband-analysis}}, and we regard the small flux ratio MCMC error as significantly underestimated.}
\end{deluxetable*}


\section{Lens-Source Relative Proper Motion} \label{sec:prop-motion}
The 2019 Keck-II follow up observations were taken 9.94 years after peak magnification in 2009. The motion of the lens and source on the sky frame is the primary cause for their apparent separation, however there is also a small component that can be attributed to the orbital motion of Earth. As this effect is of order $\leq0.1$mas for a lens at a distance of $D_{L} \geq 7$kpc, we are safe to ignore this contribution in our analysis as it is much smaller than the error bars on the stellar position measurements. The lens-source relative proper motion is measured to be $\mu_{\textrm{rel},H} = (\mu_{\textrm{rel,H,E}},\mu_{\textrm{rel,H,N}}) = (6.404 \pm 0.072, 1.788 \pm 0.145)$ mas yr$^{-1}$, where `H' indications that these measurements were made in the Heliocentric reference frame, and the `E' and `N' subscripts represent the East and North directions respectively. Converting to Galactic coordinates, these proper motions are $\mu_{\textrm{rel,H,l}} = 4.670 \pm 0.132$ mas/yr and $\mu_{\textrm{rel,H,b}} = -4.734 \pm 0.095$ mas/yr.\\
\indent Light curve modeling (section \ref{sec:light-curve}) is most conveniently performed in the Geocentric 
reference frame that moves with the Earth at the time of the event peak.
Thus, we must convert between the Geocentric and Heliocentric frames by using the relation given by \cite{dong:2009b}:

\begin{equation}\label{eq:mu-rel}
\mu_{\textrm{rel,H}} = \mu_{\textrm{rel,G}} + \frac{{\nu_{\Earth}}{\pi_{\textrm{rel}}}}{AU} \ ,
\end{equation}

\noindent where $\nu_{\Earth}$ is Earth's projected velocity relative to the Sun at the time of peak magnification. For MOA-2009-BLG-319 this value is $\nu_{\Earth \textrm{E,N}} = (29.289, 0.347)$ km/sec = $(6.175, 0.073)$ AU yr$^{-1}$ at HJD$' = 5006.99$. With this information and the relative parallax relation $\pi_{\rm{rel}} \equiv 1/D_{L} - 1/D_{S}$, we can rewrite equation \ref{eq:mu-rel} in a more convenient form:

\begin{equation}
    \mu_{\textrm{rel,G}} = \mu_{\textrm{rel,H}} - (6.175, 0.073) \times (1/D_{L} - 1/D_{S}),
\end{equation}

\noindent since we have directly calculated $\mu_{\textrm{rel,H}}$ from Keck. We use this relation in our Bayesian
analysis of the light curve, with Galactic model and Keck constraints to determine the relative proper motion in the geocentric frame of $\mu_{\textrm{rel,G}} = 6.47 \pm 0.12\,$mas. This can be compared to the value determined from the light curve MCMC without the Keck constraints of $\mu_{\textrm{rel,G}} = 6.51 \pm 0.59\,$mas, so the light curve prediction is confirmed.

\begin{figure}
\includegraphics[width=\linewidth]{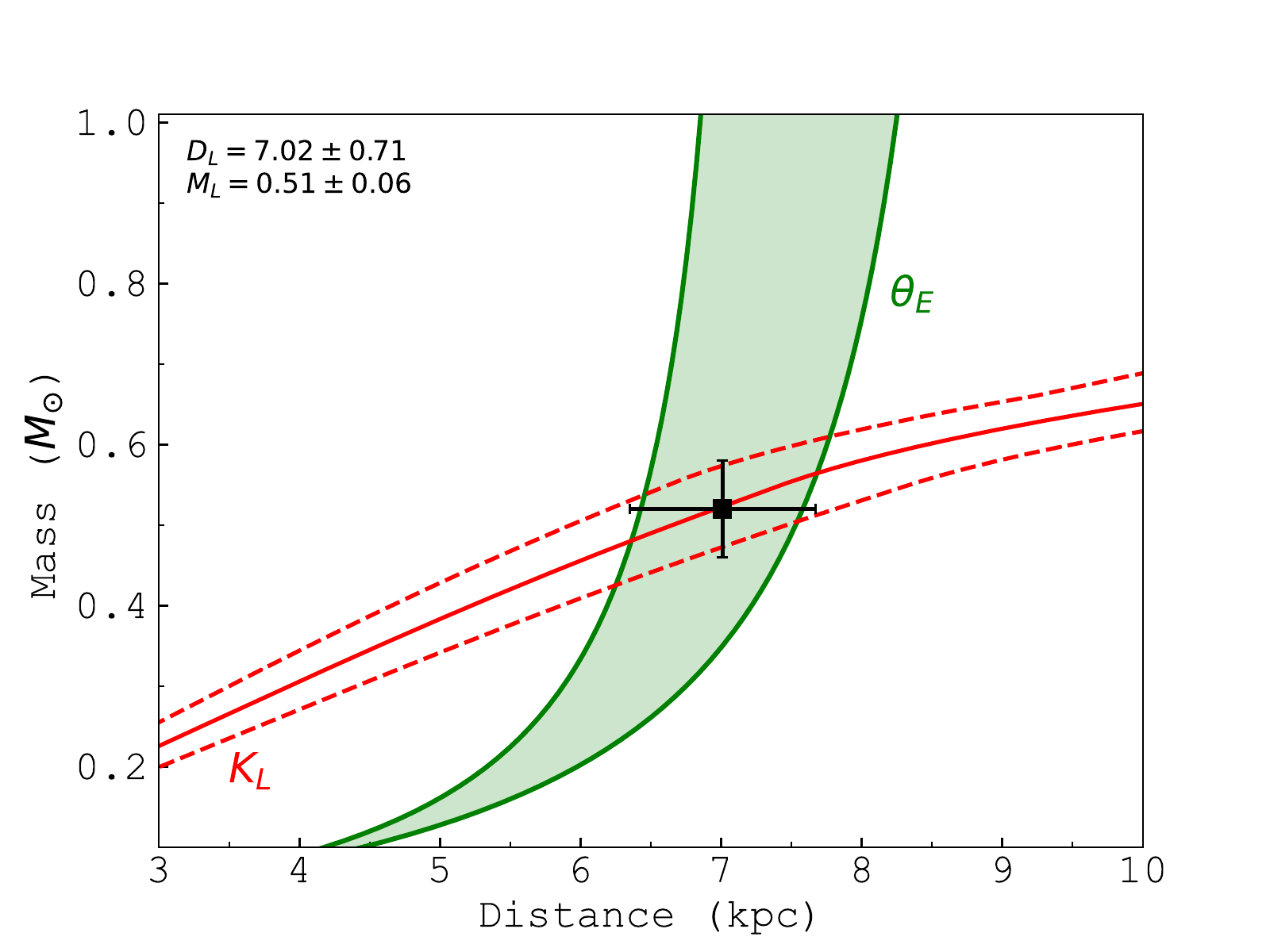}
\centering
\caption{\footnotesize Mass-Distance relation for MOA-2009-BLG-319 with constraints from the $K$ band lens flux measurement (red curve) and angular Einstein radius measurement (green curve). \label{fig:mass-dist}}
\end{figure}

\section{Lens System Properties} \label{sec:lens-properties}
The measurement of the angular Einstein radius allows us to use a mass-distance relation if we assume the distance to the source is known \citep{bennett:2008a, gaudi:2012a}:

\begin{equation}
    M_{L} = \frac{c^2}{4G}\theta_{E}\frac{D_{S}D_{L}}{D_{S}-D_{L}},
\end{equation}

\noindent where $M_{L}$ is the lens mass, $G$ and $c$ are the gravitational constant and speed of light. $D_{L}$ and $D_{S}$ are the distance to the lens and source, respectively. Figure \ref{fig:mass-dist} shows the mass-distance plane with our new direct calculation for the lens mass and distance (black). The red curve represents the constraint from the mass-luminosity relation, with dashed lines representing the error from the Keck lens flux measurement. Additionally the $\theta_{E}$ constraint is shown in green with errors dominated by the source distance uncertainty. \\
\indent As discussed in Section~\ref{sec:light-curve}, our improved photometry and improved parameterization of Galactic bulge red clump stars yields smaller $\theta_*$, $\theta_E$, and $\mu_{\textrm{rel,G}}$ values.
Our results from the re-analyzed light curve with de-trended MOA data shows a slightly fainter source star compared to \cite{miyake:2011a}. This yields a smaller angular Einstein radius and $\mu_{\textrm{rel}}$ that match the measured value better than the \cite{miyake:2011a} value.\\
\begin{figure*}[!ht]
\includegraphics[width=6.0in]{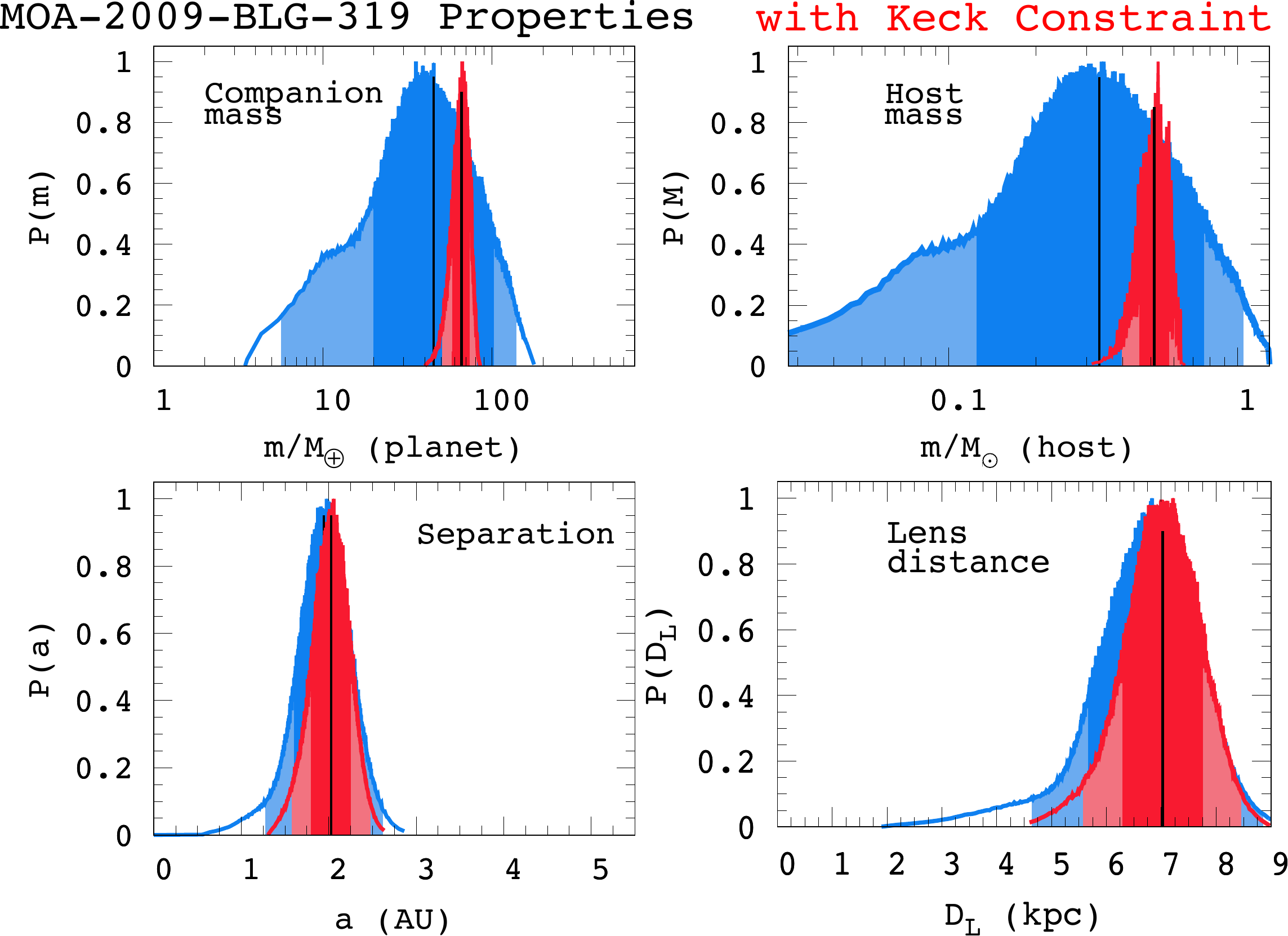}
\centering
\caption{\footnotesize The Bayesian posterior probability distributions for the planetary companion mass, host mass, their separation and the distance to the lens system are shown with only light curve constraints in blue and with the additional constraints from our Keck follow-up observations in red. The central 68.3\% of the distributions are shaded in darker colors (dark red and dark blue) and the remaining central 95.4\% of the distributions are shaded in lighter colors. The vertical black line marks the median of the probability distribution for the respective parameters. \label{fig:quad-panel}}
\end{figure*}
\indent Table \ref{tab:planet-params} shows the final planetary system results of our Bayesian analysis of the 
MCMC light curve distribution constraints from our Keck observations, as well as a Galactic model. We find that the M-dwarf lens star has a mass $M_{L} = 0.52 \pm 0.05M_{\Sun}$, with a sub-Saturn planetary companion of mass $m_{P} = 67.3 \pm 6.2M_{\Earth}$. We can calculate this planet's semi-major axis using:

\begin{equation}
    r_{\perp} = sD_{L}\theta_{E},
\end{equation}

\noindent where $s$ is the projected separation from the light curve modeling, thus we find a separation of $r_{\perp} = 2.03 \pm 0.21$AU. Additionally, the lens system is determined to be at a distance of $7.05 \pm 0.71$ kpc, very likely located in the Galactic bulge. Figure \ref{fig:quad-panel} shows the results for the physical parameters of the lens system with (red) and without (blue) the Keck constraints. The host mass and planetary mass results show very significant improvement over the unconstrained analysis, the projected separation shows marginal improvement, and the uncertainty in the lens distance is clearly still dominated by the uncertainty in the source distance as they are highly correlated.

\begin{deluxetable*}{lccc}[!htp]
\deluxetablecaption{Planetary System Properties from Lens Flux Constraints\label{tab:planet-params}}
\tablecolumns{4}
\setlength{\tabcolsep}{17.5pt}
\tablewidth{\columnwidth}
\tablehead{
\colhead{\hspace{-6cm}Parameter} & \colhead{Units} &
\colhead{Values \& RMS} & \colhead{2-$\sigma$ range}
}
\startdata
Angular Einstein Radius ($\theta_E$) & mas & $0.296 \pm 0.006$ & $0.283-0.309$\\
Geocentric lens-source relative proper motion ($\mu_{\textrm{rel,G}}$) & mas/yr & $6.472 \pm 0.121$ & $6.230-6.714$\\
Host mass ($M_{\rm host}$) & $M_{\Sun}$ & $0.524 \pm 0.048$ & $0.428-0.621$\\
Planet mass ($M_{\rm p}$) & $M_{\Earth}$ &$67.3 \pm 6.2$ & $49.8-82.2$\\
2D Separation ($a_{\perp}$) & AU & $2.03 \pm 0.21$ & $1.60-2.46$\\
3D Separation ($a_{3\textrm{d}}$) & AU & $2.90^{+1.44}_{-0.50}$ & $1.88-5.78$\\
Lens Distance (D$_{L}$) & kpc & $7.05 \pm 0.71$ & $5.60-8.45$\\
Source Distance (D$_{S}$) & kpc & $8.25 \pm 0.86$ & $6.53-9.97$\\
\enddata
\end{deluxetable*}


\section{Discussion and Conclusion} \label{sec:conclusion}

\indent Our follow-up high resolution observations of the microlensing target MOA-2009-BLG-319 have allowed us to make a direct measurement of lens flux from the host star as well as a precise determination of the direction and amplitude of the lens-source relative proper motion. Further analysis enabled us to calculate a direct mass for the star and its planetary companion. We added a novel MCMC routine to DAOPHOT\footnote[10]{\href{https://github.com/skterry/DAOPHOT-MCMC}{\scriptsize https://github.com/skterry/DAOPHOT-MCMC}} in order to retrieve precise astrometric and flux fits for the blended source and lens stars. It also allows constraints from the microlensing light curve modeling to be imposed on the analysis of high angular resolution follow-up images. Following \cite{bhattacharya:2020a}, we performed a jackknife analysis of the Keck follow-up observations because it is able to estimate uncertainties due to variations in the Keck PSF shape in multiple images. We used these jackknife error bars for our final analysis. These methods provide more accurate results than previously used techniques for crowded field photometry in AO imaging. These routines can be used in future analyses of highly blended microlensing follow-up targets, and eventually, can form the basis for the \textit{Roman} mass measurement method.\\
\indent The MOA-2009-BLG-319 microlensing event has a planet-to-star mass ratio of $q = (3.856\pm 0.029)\times 10^{-4}$, which puts it in the range of the mass ratio desert originally predicted by \citet{ida:2004a} and confirmed more recently
by \citet{suzuki:2018a}. This prediction was based on the runaway gas accretion scenario that has been considered a standard part of the core accretion theory \citep{pollack:1996aa}, but is based on a one-dimensional calculation. The \citet{suzuki:2018a} analysis found a discrepancy between the planet mass ratio distribution found by microlensing and this predicted mass ratio gap, at $10^{-4} < q < 4\times 10^{-4}$, thought to be caused by the rapid ``runaway" growth. It was thought to be unlikely that planet growth would terminate during this predicted very rapid growth phase. But, the microlensing results of \citet{suzuki:2016a} show no evidence of this predicted gap.\\
\indent One possible explanation for this contradiction might be that the runaway gas accretion phase only occurs for
stars of approximately solar type, which was the original focus of the core accretion theory, while microlensing 
probes not only solar type stars, but also lower mass stars and even stellar remnants. Our high angular resolution follow-up observations can test this possibility by measuring host star masses for the 7 events of the 30 in the \citet{suzuki:2016a} sample that fall in the mass ratio range $10^{-4} < q < 4\times 10^{-4}$. Mass measurements have previously been made for two of the 7 \citet{suzuki:2016a} host stars with planets in this range. \citet{bhattacharya:2018a} has measured a host mass of $M_{\rm host} = 0.58\pm 0.05 M_\odot$ and a planet mass of $m_p = 39\pm 9 M_\oplus$ for planetary microlensing event OGLE-2012-BLG-0950, and \citet{bennett16} have measured host and planet masses of $M_{\rm hosts} = 0.71\pm 0.12 M_\odot$  and
$m_p = 80\pm 13 M_\oplus$ for the OGLE-2007-BLG-349L lens system, although in this case the host is a close binary pair of $0.41 \pm 0.07 M_\odot$ and $0.31 \pm 0.07 M_\odot$ in a $\sim${}$10$ day orbit. Our group has also measured the mass of a more massive host star, OGLE-2012-BLG-0026L \citep{Beaulieu:2016a}, with $M_{\rm host} = 1.06\pm 0.05 M_\odot$, with one planet in the mass ratio range of the predicted gap, $10^{-4} < q < 4\times 10^{-4}$. The sub-Saturn planet has a mass of $46\pm 2 M_\oplus$, and it is accompanied by a more massive planet with a mass of $265\pm 20 M_\oplus$. However, this event is not in the \citet{suzuki:2016a} statistical sample.\\
\indent The addition of the MOA-2009-BLG-319L system to this collection with host and planet masses of $M_{\rm host} =0.52 \pm 0.05 M_\odot $ and $m_{p} = 66 \pm 8M_{\Earth}$ continues the trend of finding host masses within a factor of two of a solar mass, and this suggests that the lack of this mass ratio gap at $10^{-4} < q < 4\times 10^{-4}$ is not caused by some dramatic change in the mass ratio for host stars with very low masses. Such a conclusion would be supported by the theoretical work of Szul{\'a}gyi et al. (in preparation), who show that the runaway gas accretion phase is likely to be terminated very quickly by the formation of a circumplanetary disk, which can result in many planets in the predicted gap. Further results from our high angular resolution follow-up imaging program will provide a stronger test of these core accretion processes, with additional mass measurements for the host stars of sub-Saturn mass planets orbiting beyond the snow line. A more definitive answer to this and other questions regarding the demographics of planets in wider orbits will come from the RGES, which will have high enough angular resolution so that follow-up observations will not be needed for the majority of exoplanets discovered.

The authors thank Dr. Peter Stetson for advice on modifications to the DAOPHOT-II software. We also thank the anonymous referee for constructive comments that led to a stronger manuscript. This work was performed in part under contract with the Center for Research and Exploration in Space Sciences and Technologies (CRESST-II). The Keck observations were supported by a NASA Keck PI Data Award, 80NSSC18K0793, administered by the NASA Exoplanet Science Institute. Data presented herein were obtained at the W. M. Keck Observatory from telescope time allocated to the National Aeronautics and Space Administration through the agency's scientific partnership with the California Institute of Technology and the University of California. The Observatory was made possible by the generous financial support of the W. M. Keck Foundation. The authors wish to recognize and acknowledge the very significant cultural role and reverence that the summit of Maunakea has always had within the indigenous Hawaiian community. We are most fortunate to have the opportunity to conduct observations from this mountain. DPB, AB, and CR were supported by NASA through grant NASA-80NSSC18K0274. This work was supported by the University of Tasmania through the UTAS Foundation and the endowed Warren Chair in Astronomy and the ANR COLD-WORLDS (ANR-18-CE31-0002). Work by NK is supported by JSPS KAKENHI Grant Number JP18J00897. This research was also supported in part by the Australian Government through the Australian Research Council Discovery Program (project number 200101909) grant awarded to Cole and Beaulieu. This work made use of data from the Astro Data Lab at NSF's OIR Lab, which is operated by the Association of Universities for Research in Astronomy (AURA), Inc. under a cooperative agreement with the National Science Foundation. Some of this research has made use of the NASA Exoplanet Archive, which is operated by the California Institute of Technology, under the Exoplanet Exploration Program.

\textit{Software}: DAOPHOT-II \citep{stetson:1987a}, DAOPHOT-MCMC (this work), gnuplot, Matplotlib \citep{hunter:2007a}, Numpy \citep{oliphant:2006a}.

\bibliographystyle{aasjournal}
\bibliography{Terry_moa09319.bib}

\end{document}